\documentclass[conference,10pt]{IEEEtran} 

\usepackage{graphicx}
\usepackage{color}
\usepackage{placeins}
\usepackage{float}
\usepackage{tabularx,colortbl}
\usepackage{amssymb}
\usepackage{amsthm}
\usepackage{cite}
\usepackage{amsmath}
\usepackage{algorithm}
\usepackage[noend]{algpseudocode}

\def\Htran{\mbox{\tiny $\mathrm{H}$}}

\newcommand{\fracSum}[1]{{\underset{{#1}}{\sum}}}

\newcommand{\vect}[1]{\mathbf{#1}}

\newcommand{\br}[1]{{\left\{ #1 \right\}}}

\theoremstyle{plain}

\newtheorem{theorem}{Theorem}

\newtheorem{definition}{Definition}

\IEEEoverridecommandlockouts

\begin{document}

\title{\huge Pilot Clustering in Asymmetric Massive MIMO Networks
\thanks{\copyright~2015 IEEE. Personal use of this material is permitted. Permission from IEEE must be obtained for all other uses, in any current or future media, including reprinting/republishing this material for advertising or promotional purposes, creating new collective works, for resale or redistribution to servers or lists, or reuse of any copyrighted component of this work in other works.}
}%
\author{ 
\authorblockN{Rami Mochaourab$^*$, Emil Bj\"ornson$^\ddagger$, and Mats Bengtsson$^*$ \thanks{This research has received funding from ELLIIT and CENIIT.}}
\authorblockA{$^*$ Signal Processing Department, ACCESS Linnaeus Centre, KTH Royal Institute of Technology, Sweden\\ 
$^\ddagger$ Department of Electrical Engineering (ISY), Link\"oping University, Sweden}}

\maketitle


\begin{abstract}
We consider the uplink of a cellular massive MIMO network. 
Since the spectral efficiency of these networks is limited by pilot contamination, the pilot allocation across cells is of paramount importance. However, finding efficient pilot reuse patterns is non-trivial especially in practical asymmetric base station deployments. In this paper, we approach this problem using coalitional game theory. Each cell has its own unique pilots and can form coalitions with other cells to gain access to more pilots. We develop a low-complexity distributed algorithm and prove convergence to an individually stable coalition structure. Simulations reveal fast algorithmic convergence and substantial performance gains over one-cell coalitions and full pilot reuse.
\end{abstract}

%\vspace{-1mm}

\IEEEpeerreviewmaketitle

\section{Introduction}\label{sec:introduction}%
Massive MIMO (multiple-input, multiple-output) technology has over the last few years emerged from a theoretical concept \cite{Marzetta2010a} to a key solution for future wireless networks \cite{Boccardi2014a}. This is because it can improve the sum spectral efficiency (bit/s/Hz/cell) of cellular networks by orders of magnitude \cite{Bjornson2016a}, without the need for more spectrum or more base stations (BSs). In massive MIMO, each BS is equipped with an array of hundreds of  active antennas, which are processed coherently to improve the signal quality in the uplink and downlink \cite{Rusek2013a}.

Massive MIMO systems require channel state information (CSI) at the BSs, for example, to separate  uplink signals sent in parallel by different user equipments (UEs). CSI is acquired from uplink pilot signaling. The pilot sequences are precious resources in cellular networks since accurate CSI estimation requires low interference during pilot transmission (i.e., low so-called pilot contamination \cite{Jose2011b}). Contemporary networks have an over-provision of pilots---many more orthogonal pilots than active UEs per cell---thus pilot contamination is essentially alleviated by allocating the pilots at random. In contrast, massive MIMO networks attempt to schedule as many users as possible to achieve a high sum spectral efficiency (SE) \cite{Bjornson2016a}.

An efficient and robust way to mitigate pilot contamination is fractional pilot reuse, where only a subset of the pilot resources are utilized in each cell \cite{Yang2013a,Li2012a,Bjornson2016a}. This is conceptually simple in symmetric networks (e.g., one-dimensional cases as in \cite{Li2012a} or two-dimensional cases with hexagonal cells as in \cite{Yang2013a,Bjornson2016a}); one can color the cells in a symmetric pattern and divide the pilots so that only cells with the same color use the same subset. The pilot clustering in practical asymmetric deployments, where every cell has a unique shape, is non-trivial and must be optimized for each particular deployment. The purpose of this paper is to use game theory to develop an algorithm for adaptive pilot clustering, which can be applied in cellular networks with arbitrary asymmetric cell geometries.

Coalitional game models offer structured mechanisms to find cooperation between decision makers (players). A characteristic of these mechanisms is their natural implementability in a distributed way. With such merits, coalitional game theory has found many applications in communication networks; see the tutorial \cite{Saad2009a}. A class of coalitional games, called the partition form \cite{Thrall1963}, takes into account the dependencies of the players' utility functions on the overall partition of players into distinct coalitions, called coalition structure. The application of such game models to determine which BSs (players) cooperate with each other is appropriate in this paper as the performance of each cell depends on the whole coalition structure. 

The solution of a coalitional game in partition form is a coalition structure with specified stability requirements. The stability requirements are generally related to the feasible transitions from one coalition structure to another. One model of stability, called individual stability, restricts a single player to move from one coalition to another \cite{Bogomolnaia2002}. Such models, which we utilize in this paper, have been applied for channel sensing and access in cognitive radio \cite{Saad2012} and  for coalition formation in the MIMO interference channel \cite{Zhou2013}.

In this paper, we assume that each BS has a set of unique pilot sequences. We propose a distributed mechanism to find the sets of BSs that cooperate to gain access to each others' pilots and thereby serve more UEs. A set of BSs have the incentive to cooperate whenever the larger number of UEs that can be scheduled in their cells leads to an improvement in their individual per-cell average SE.
Our mechanism is a coalition formation algorithm which converges to an individually stable coalition structure under what we define as searching budget constraints. In doing so, the complexity of the algorithm is controlled. Simulation results reveal fast convergence of the algorithm giving considerable performance gains over one-cell coalitions and full pilot reuse.%
\section{System Model \& Sum Spectral Efficiency}
\label{sec:system-model}

We consider the uplink of a cellular massive MIMO network with $L$ cells, each associated with an index in the set $\mathcal{L} = \{1, \ldots,L\}$. BS $j$ is equipped with an array of $M$ antennas and serves $K_j$ UEs. The transmission is divided into frames of $T_c$ seconds and $W_c$ Hz, such that channel between each UE and BS has a constant channel response within a frame, but is different between frames. Consequently, each frame contains $S = T_c W_c$ transmission symbols.
In each uplink frame, $B$ symbols are allocated for pilot signaling and the remaining $S-B$ symbols are used for uplink data. The $B$ pilot symbols permit $B$ orthogonal pilot sequences; that is, only $B$ UEs in the entire network can transmit pilots without interfering with each other. In this paper, we study how the $L$ cells can share these pilots to maximize the average sum SE. Since pilot contamination is mainly a problem in highly loaded networks, we assume that there is at least $B$ potential UEs per cell---it is up to each BS to decide how many of them that are active.

Each cell is given a fraction $\frac{B}{L}$ of unique pilot sequences, where $\frac{B}{L}$ for convenience is assumed to be an integer. BS $j$ can keep its $\frac{B}{L}$ pilots by itself and serve $K_j = \frac{B}{L}$ UEs without any pilot contamination. Alternatively, it can form a coalition with other cells to share the access to each others' pilots. 

\begin{definition}[Coalition structure] \label{def:coalition-structure}
A coalition structure $\mathcal{C}$ is a partition of $\mathcal{L}$, the grand coalition, into a set of disjoint coalitions $\{\mathcal{S}_1,\ldots,\mathcal{S}_N\}$ with $\bigcup_{n = 1}^{N} \mathcal{S}_n \!=\! \mathcal{L}$ and $\bigcap_{n= 1}^{N} \mathcal{S}_n \!=\! \emptyset$.
\end{definition}

For example, let $\Phi_j (\mathcal{C})$ denote the coalition in $\mathcal{C}$ that BS $j$ belongs to. The coalition members have access to $\frac{B}{L} | \Phi_j (\mathcal{C}) | $ pilots, where $| \cdot |$ denotes the cardinality of a set. Then, BS $j$ can serve $K_j = \frac{B}{L} | \Phi_j (\mathcal{C}) |$ UEs, but the drawback is that the cells in the coalition contaminate each others' pilot transmissions.

\begin{figure}[t]
  \centering
  \includegraphics[width=\linewidth,clip]{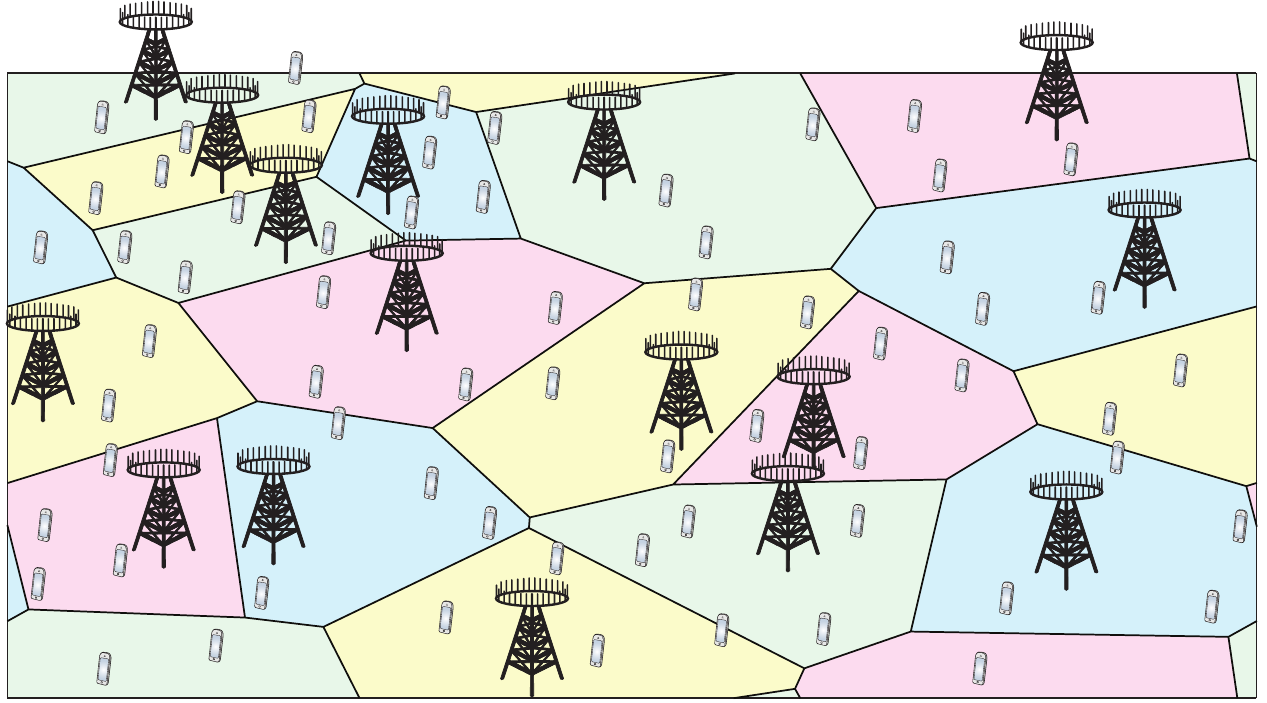}
  \caption{\label{fig:systemmodel}  Illustration of a massive MIMO network with $L$ cells and wrap-around. Each cell contains a BS with $M$ antennas and the colors indicate clusters of cells that use the same pilots.} \vspace{-4mm}
\end{figure}

Fig.~\ref{fig:systemmodel} gives an example of a cellular network with $L=16$ cells in an area with vertical and horizontal wrap-around. The cells have formed four coalitions: green, yellow, red, and blue. Since each coalition has four members, each BS has access to $\frac{B}{4}$ pilots and serves $\frac{B}{4}$ UEs in each frame. Pilot contamination is only caused between cells with the same color. 

%\begin{figure}[t]
%  % Requires \usepackage{graphicx}
%  \centering
%  \includegraphics[width=\linewidth,clip]{illustrations/protocol}
%  \caption{\label{fig:protocol} Frame structure in the uplink of a massive MIMO system, where $S = T_c W_c$ is the number of symbols per frame.}
%\end{figure}

\subsection{Average Uplink Spectral Efficiency}

The coalition formation is based on maximizing the SE in each cell. The vast majority of SE expressions for massive MIMO assumes that each BS serves the same number of UEs (cf.~\cite{Marzetta2010a,Hoydis2013a,Bjornson2016a}). In contrast, the BSs may form coalitions of different sizes and thus serve unequal numbers of UEs herein. We now generalize the SE expressions from \cite{Bjornson2016a} to handle this.

The UEs are randomly distributed in the serving cell. In a certain frame, suppose that $\vect{z}_{lk} \in \mathbb{R}^2$ is the position of the $k$th UE in cell $l$. The channel response $\vect{h}_{jlk} \in \mathbb{C}^M$ between this UE and BS $j$ is modeled as Rayleigh fading:
\begin{equation} \label{eq:channel-distribution}
\vect{h}_{jlk} \sim \mathcal{CN}\Big(\vect{0},d_j(\vect{z}_{lk}) \vect{I}_M \Big),
\end{equation}
where $\vect{I}_M$ is the $M \times M$ identity matrix. The function $d_j(\vect{z})$ gives the variance of the channel attenuation from a UE position $\vect{z}$ to BS $j$. The value of $d_j(\vect{z}_{lk})$ is known at BS $j$. The UEs apply power control with the purpose of achieving uniform performance and avoid near-far blockage. More precisely, the UE at location $\vect{z}_{jk}$ uses a transmit power of $\rho/ d_j(\vect{z}_{jk})$ per symbol, where $\rho$ is a design parameter. The resulting signal-to-noise ratio (SNR) at any antenna at the serving BS is $\rho/\sigma^2$, where $\sigma^2$ is the noise variance.

Orthogonal pilots are used within each cell, while the $k$th UE of each cell in a coalition uses the same pilot. By correlating the received signals with corresponding pilot sequences \cite{Hoydis2013a}, the received pilot signal $\vect{y}_{jk}^{\mathrm{pilot}}\in \mathbb{C}^{M}$ at BS $j$ for its UE $k$ is 
\begin{equation} \label{eq:system-model-pilot}
\vect{y}_{jk}^{\mathrm{pilot}} = \sum_{l \in \Phi_j (\mathcal{C}) }  \sqrt{ \frac{\rho}{d_l(\vect{z}_{lk})} B } \vect{h}_{jlk} + \boldsymbol{\eta}_{jk}, ~~ k = 1,\ldots,K_j,
\end{equation}
where $\boldsymbol{\eta}_{jk} \sim \mathcal{CN}(\vect{0},\sigma^2 \vect{I}_M)$ is additive noise. Notice that there is only interference from one UE per cell in the coalition $\Phi_j (\mathcal{C}) $, and no interference from other coalitions.
Based on $\vect{y}_{jk}^{\mathrm{pilot}}$,  BS $j$ computes the MMSE estimate of $\vect{h}_{jjk}$ \cite{Bjornson2016a}:
\begin{equation} \label{eq:LMMSE-estimator}
  \hat{\vect{h}}_{jjk} =  \frac{ \sqrt{ \rho d_j(\vect{z}_{jk})B }   }{ \sum_{\ell \in \Phi_j (\mathcal{C})}   \frac{\rho d_j(\vect{z}_{\ell k}) }{d_\ell(\vect{z}_{\ell k})} B + \sigma^2 } \vect{y}_{jk}^{\mathrm{pilot}}.
\end{equation}

During uplink data transmission, all BSs are active and the received signal $\vect{y}_j \in \mathbb{C}^{M}$ at BS $j$ is 
\begin{equation} \label{eq:system-model}
\vect{y}_j^{\mathrm{data}} = \sum_{l = 1}^{L} \sum_{k=1}^{K_l} \sqrt{ \frac{\rho}{d_l(\vect{z}_{lk})} }  \vect{h}_{jlk} x_{lk} + \vect{n}_{j},
\end{equation}
where $x_{lk} \in \mathbb{C}$ is the symbol transmitted by UE $k$ in cell $l$. This signal is normalized as $\mathbb{E}\{ | x_{lk} |^2 \} = 1$, while the corresponding UL transmit power is $ \frac{\rho}{d_l(\vect{z}_{lk})}$, as defined earlier. The additive noise is modeled as $\vect{n}_{j} \sim \mathcal{CN}(\vect{0},\sigma^2 \vect{I}_M)$.

Linear receive combining is used in massive MIMO to separate each UE's signal from the interfering signals. BS $j$ selects a combining vector $\vect{g}_{jk} \in \mathbb{C}^{M}$ for each of its $K_j$ UEs and multiply it with the received signals in \eqref{eq:system-model}, as $\vect{g}_{jk}^{\Htran} \vect{y}_j^{\mathrm{data}} $, to obtain an effective scalar signal. Let $\hat{\vect{H}}_{j} = [\hat{\vect{h}}_{jj1} \, \ldots \, \hat{\vect{h}}_{jj K_j}] \in \mathbb{C}^{M}$
be the estimated channels in cell $j$. Two typical combining schemes are maximum ratio combining (MRC), which obtains the highest signal gain by setting
\begin{equation}
[\vect{g}_{j 1}^{\mathrm{MRC}} \, \ldots \,\, \vect{g}_{j K_j}^{\mathrm{MRC}}  ] =   \hat{\vect{H}}_{j} ,
\end{equation}
and zero-forcing combining (ZFC) where the pseudo-inverse of $ \hat{\vect{H}}_{j}$ is used to suppress intra-cell interference:
\begin{equation}
[\vect{g}_{j 1}^{\mathrm{ZFC}} \, \ldots \,\, \vect{g}_{j K_j}^{\mathrm{ZFC}}  ] =  (  \hat{\vect{H}}_{j}^{\Htran}  \hat{\vect{H}}_{j}  )^{-1} \hat{\vect{H}}_{j}.
\end{equation}

\begin{figure*}[ht!]
\begin{align}
I_{j}^{\mathrm{MRC}}(\mathcal{C}) &= \fracSum{l \in \Phi_j (\mathcal{C}) \setminus \{ j \}} \! \Bigg( \mu^{(2)}_{jl} + \frac{\mu^{(2)}_{jl}-\left( \mu^{(1)}_{jl} \right)^2}{M} \Bigg) \!+\! \Bigg( 
\sum_{\mathcal{S} \in \mathcal{C} } \sum_{l \in \mathcal{S}} \mu_{jl}^{(1)} \frac{ B | \mathcal{S} | }{ L }  + \frac{\sigma^2}{   \rho } \Bigg) \!\! \Bigg( \frac{ \fracSum{l \in \Phi_j (\mathcal{C})} \mu_{jl}^{(1)}  + \frac{\sigma^2}{ B \rho} }{M}\Bigg)    \label{eq:achievable-SINR-MR2} \\
I_{j}^{\mathrm{ZFC}}(\mathcal{C}) &= \fracSum{l \in \Phi_j (\mathcal{C})  \setminus \{ j \}} \! \Bigg( \mu^{(2)}_{jl} + \frac{\mu^{(2)}_{jl}-\left( \mu^{(1)}_{jl} \right)^2}{M -  \frac{B | \Phi_j (\mathcal{C}) | }{L} } \Bigg) \! +\!  \left( \sum_{\mathcal{S} \in \mathcal{C} } \sum_{l \in \mathcal{S}} \mu_{jl}^{(1)} \frac{ B | \mathcal{S} | }{ L}  \!-\! 
\frac{ \fracSum{l \in \Phi_j (\mathcal{C})  }  (\mu_{j l}^{(1)} )^2 \frac{ B | \Phi_j (\mathcal{C}) | }{ L} }{  \fracSum{\ell \in \Phi_j (\mathcal{C}) } \mu_{j \ell}^{(1)} + \frac{\sigma^2}{ B \rho}   }
 +  \frac{\sigma^2}{ \rho } \! \right)\! \!\! \Bigg( \! \frac{ \fracSum{l \in \Phi_j (\mathcal{C}) } \mu_{jl}^{(1)}  + \frac{\sigma^2}{ B \rho} }{M-  \frac{B | \Phi_j (\mathcal{C}) |}{L} } \Bigg)  \label{eq:achievable-SINR-ZF2}
\end{align} \vskip-2mm
\hrulefill
\vskip-3mm
\end{figure*}

The following theorem provides an average sum SE in the cells with MRC and ZFC, and generalizes previous results in \cite{Bjornson2016a} for equal number of UEs per cell.

\begin{theorem} \label{theorem:sum-spectral-efficiency}
For a given coalition structure $\mathcal{C}$, a lower bound on the average ergodic sum capacity in cell $j$ is
\begin{equation} \label{eq:utility-functions}
U_j(\mathcal{C}) = \left( 1 - \frac{B}{S} \right) \frac{B}{L} | \Phi_j (\mathcal{C}) | \,\log_2 \left(  1 + \frac{1}{I_{j}^{\mathrm{scheme}}(\mathcal{C}) } \right),
\end{equation}
where the interference term $I_{j}^{\mathrm{scheme}}$ is given in \eqref{eq:achievable-SINR-MR2}  for MRC and in \eqref{eq:achievable-SINR-ZF2} for ZFC (both at the top of the page).
The following propagation parameters appear in these expressions:
\begin{align} \label{eq:mu-definition1}
\mu^{(\gamma)}_{jl} &= \mathbb{E}_{\vect{z}_{lm}} \left\{ \left( \frac{d_j(\vect{z}_{lm}) }{ d_l(\vect{z}_{lm})} \right)^{\gamma} \right\} \quad \textrm{for} \,\, \gamma = 1,2. 
\end{align}
\end{theorem}
\begin{IEEEproof}
The proof is along the lines of the proofs of Lemma 2 and Th.~1 in 
\cite{Bjornson2016a} and are omitted for breviety.
\end{IEEEproof}

The factor $( 1 - \frac{B}{S} ) $ in \eqref{eq:utility-functions} is the pilot signaling overhead, while second factor is the number of active UEs in cell $j$: $K_j = \frac{B}{L} | \Phi_j (\mathcal{C}) |$. The interference terms $I_{j}^{\mathrm{MRC}}(\mathcal{C}) $ and $I_{j}^{\mathrm{ZFC}}(\mathcal{C})$ have intuitive interpretations. The first part in both expressions is the pilot contamination and is only impacted by the cells that have formed a coalition with BS $j$. The second part is the conventional inter-user interference (from all cells). MRC suppresses this part by the full array gain of $M$, while 
ZFC cancels out part of the interference and suppresses the remaining interference by a reduced array gain of $M-K_j$.

The parameter $\mu^{(1)}_{jl}$ in \eqref{eq:mu-definition1} is the average ratio between the channel variance to BS $j$ and the channel variance to BS $l$, for a UE in cell $l$. Along with its second-order moments, $\mu^{(2)}_{jl}$, these parameters characterize the network topology. Notice that $\mu^{(1)}_{jj} = \mu^{(2)}_{jj} =1$, while the values get smaller as further cell $j$ and cell $l$ are apart. In general, $\mu^{(1)}_{jl} \neq \mu^{(1)}_{lj}$ for $j \neq l$.
% In general, we have $\mu^{(1)}_{jl} \neq \mu^{(1)}_{lj}$ while equality only holds for symmetric networks where all cells have the same shape.
 
The average sum SE $U_j(\mathcal{C})$ in \eqref{eq:utility-functions}, the utility function of BS $j$ in the remainder of this paper, should preferably be as large as possible\footnote{Note that maximizing the sum SE might lead to operating points with many active users and low SE per user, but this is still beneficial for all users as compared to time-sharing where each user is only active part of time but exhibit a higher SE when being active.}. There are thus $L$ utilities which depend on the combining scheme (e.g., MRC or ZFC) and on the coalition structure $\mathcal{C}$. Since the number of possible coalition structures equals the $L$th Bell number, which has a faster growth than exponential with $L$, finding a globally optimal pilot assignment is hard. Therefore, we formulate next the design problem as a coalitional game to provide a distributed and efficient algorithm.%

\section{Coalitional Game}\label{sec:coalition_formation}
Cooperation between the BSs can be analyzed using coalitional games \cite{Osborne1994}. Since the average SE of each cell, given in Theorem~\ref{theorem:sum-spectral-efficiency}, depends on the coalition structure (Definition \ref{def:coalition-structure}), we need to study the coalitional game in partition form \cite{Thrall1963}, which we formulate by the tuple $\langle \mathcal{L}, (q_j)_{j\in \mathcal{L}}, (\tilde{U}_j)_{j\in \mathcal{L}} \rangle$. Here, the set of players corresponds to the set of BSs $\mathcal{L}$. Each player in $\mathcal{L}$ is endowed with \emph{searching budget} $q_k\in\mathbb{N}$ which limits the number of searches he can perform to find a coalition to join. The utility of a player $j$ is formulated to be
\begin{equation} \label{eq:restricted_utility}
\tilde{U}_j(\mathcal{C},\eta_j) = \begin{cases}
   U_j(\mathcal{C}) &\text{if } \eta_j \leq q_j  \\
   0       &\text{otherwise}\\
  \end{cases},
\end{equation}
which is specified by its SE ${U}_j$ whenever a player $j$ has not exhausted its searching budget $q_j$ where $\eta_j \in\mathbb{N}$ represents the number of searches player $j$ has already performed.

From Theorem~\ref{theorem:sum-spectral-efficiency}, the utility of cell $j$ depends on which members are in its coalition $\Phi_j(\mathcal{C})$ through the pilot contamination term as well as the interference term determined by the structure of the coalitions forming outside $\Phi_j(\mathcal{C})$. Therefore, so-called \emph{externalities} exist. Specifically, our game belongs to the category of negative externalities, since the merging of coalitions reduces the utility of all coalitions not involved in the merging due to the increased number of scheduled users and thereby the increased interference.%\cite{Yi1997}

We adopt the game theoretic assumptions which implies that each player's behavior follows the maximization of his utility based on the discovery of profitable opportunities \cite{Osborne1994}. Such behavior is important for distributed coalition formation which we specify and discuss next.

\subsection{Coalition Formation}
Coalition formation represents the dynamics which lead to stable coalition structures. We use a coalition formation model from \cite{Bogomolnaia2002} in which a single player is allowed to leave its coalition and join another.%Such a deviation is defined as follows.
\begin{definition}[Deviation]\label{def:deviation}
A cell $j\in \mathcal{L}$ leaves its current coalition $\Phi_j(\mathcal{C})$ to {join} coalition $\mathcal{S} \in \mathcal{C} \cup \{\emptyset\}$. In doing so, the coalition structure $\mathcal{C}$ changes to $\mathcal{C}^{\mathcal{S}}$. We capture this change in the coalition structure by the notation $\mathcal{C}^{\mathcal{S}} \overset{j}{\longleftarrow} \mathcal{C}$.
\end{definition}

% Observe that a deviation by a player entitles a search for alternatives within the current coalition structure. Given a coalition structure $\mathcal{C}$, the number of searches by a player $j$ is upper bounded by the number of coalitions outside $\Phi_j(\mathcal{C})$ and the empty set. Accordingly, by defining a searching budget for each player $j$ as $q_j$ we can control the complexity for coalition formation. 

According to individual stability \cite{Bogomolnaia2002}, a deviation is admissible if a player strictly improves his performance by leaving its coalition to join another coalition. In addition, the members of the coalition which he joins should not reduce their utility. 
%The deviation of a player entitles a search for alternatives within the current coalition structure. Given a coalition structure $\mathcal{C}$, the number of searches is upper bounded  by the number of coalitions outside $\Phi_j(\mathcal{C})$ and the empty set. In order to control the complexity for coalition formation, we define a searching budget for each BS.

\begin{definition}[Admissible deviation]\label{def:admissable}
A deviation $\mathcal{C}^{\mathcal{S}} \overset{j}{\longleftarrow} \mathcal{C}$ is admissible if $\tilde{U}_j(\mathcal{C}^{\mathcal{S}},\eta_j)  > \tilde{U}_j(\mathcal{C},\eta_j),$ and $\tilde{U}_k(\mathcal{C}^{\mathcal{S}},\eta_k) \geq \tilde{U}_k(\mathcal{C},\eta_k)$, for all $k \in \mathcal{S}$.
\end{definition}

The requirement imposed through the admissible deviation is suitable in our setting due to the fact that each cell exclusively owns a set of pilots. Thus, any BS which wants to join a coalition by sharing its pilots with its members must first ask their permission.

%\begin{definition}[Searching budget]\label{def:budget}
%The maximum number of searches player $j$ (BS $j$) can perform in order to find a coalition to join is $q_j\in\mathbb{N}$.
%\end{definition}

Based on the player's deviation model, we utilize the following stability concept defined in \cite{Bogomolnaia2002}.

\begin{definition}[Individual stability]\label{def:Individual_stability}
A coalition structure $\mathcal{C}$ is individually stable if there exists no $j\in\mathcal{L}$ and coalition $\mathcal{S}$ such that deviation $\mathcal{C}^{\mathcal{S}} \overset{j}{\longleftarrow} \mathcal{C}$ is admissible.
\end{definition}

%\begin{definition}[Individual stability]\label{def:Individual_stability}
%A coalition structure $\mathcal{C}$ is individually stable if there do not exist $j \in \mathcal{L}$ in a coalition $\Phi_j(\mathcal{C})$ and a coalition $\mathcal{S} \in \mathcal{C} \cup \{\emptyset\}$ such that $\mathcal{C}^{\mathcal{S}} \overset{j}{\longleftarrow} \mathcal{C}$ satisfies $U_j(\mathcal{C}^{\mathcal{S}}) > U_j(\mathcal{C})$ and $U_k(\mathcal{C}^{\mathcal{S}}) \geq U_k(\mathcal{C}), \text{ for all } k \in \mathcal{S}$.
%\end{definition}

% Individual stability necessitates that no player can strictly improve its performance by leaving its current coalition to join another coalition. In addition, the members of the coalition which the deviating player joins should not reduce their utility. Such a solution concept is suitable in our setting due to the fact that each cell exclusively owns a set of pilots. Thus, any BS which wants to join a coalition by sharing its pilots with its members must first ask their permission.

% \subsection{Coalition Formation Algorithm}
%Our algorithm is based on the deviation model (Definition \ref{def:deviation}), stability concept (Definition \ref{def:Individual_stability}) and the searching budget constraint (Definition \ref{def:budget}).

The coalition formation algorithm which leads to individually stable coalition structures is described in Algorithm~\ref{alg:coalition0}. The coalition structure is initialized with singleton coalitions, corresponding to the noncooperative state in which no pilots are shared between the cells. Note that the algorithm can be initialized with any coalition structure. A BS $j$ is selected at random to check if a deviation is profitable. Based on the local knowledge of the current coalition structure $\mathcal{C}_t$ and the propagation parameters, BS $j$ can calculate its utility in~\eqref{eq:restricted_utility} if it joins other coalitions. Then, BS $j$ selects a coalition at random in which it would profit by joining (Line 3) and asks its members for permission to join (Line 4). Here, we assume that the BSs are able to communicate and exchange such application-type messages. Each BS in a coalition can calculate its utility locally for the case the asking BS enters the coalition. In Line 5, the number of searches by BS $j$ is incremented. If the deviation is admissible (Definition~\ref{def:admissable}), then it joins the coalition (Line 7) and the coalition structure changes accordingly (Line 8). Coalition formation stops when no deviations take place anymore.

\begin{theorem}
Algorithm \ref{alg:coalition0} converges to an individually stable coalition structure with an upper bound on the number of deviations as $t \leq \sum_{j \in \mathcal{L}} q_j$.
\end{theorem}
\begin{IEEEproof}
The convergence of Algorithm \ref{alg:coalition0} is guaranteed due to the searching budget incorporated in the utility function in~\eqref{eq:restricted_utility} leading to $t\leq\sum_{j \in \mathcal{L}} q_j$. The stability result follows from iterating over all deviation opportunities until individual stability (Definition~\ref{def:Individual_stability}) is satisfied.
\end{IEEEproof}

\begin{algorithm}[t]
\caption{\label{alg:coalition0} Coalition formation algorithm.}
\begin{algorithmic}[1]
%\Statex \textbf{Input}: {$\mathcal{L}$, $(q_1,\ldots, q_L)$}
\Statex \textbf{Initilize}: $t = 0$, $\mathcal{C}_{0} = \br{\br{1},\ldots,\br{K}}$, $\eta_j = 0, j \in \mathcal{L}$
\Repeat
\ForAll{BSs $j \in \mathcal{L}$}
%\Statex ~~\underline{At BS $j$:}
%\State \parbox[t]{\dimexpr\linewidth-\algorithmicindent-\algorithmicindent}{Calculate utilities $\br{U_j(\mathcal{C}^{\mathcal{S}}_{t})}_{\mathcal{S} \in \mathcal{D}_j}$ with $\mathcal{C}^{\mathcal{S}}_{t}\overset{j}{\longleftarrow} \mathcal{C}_t$ and $\mathcal{D}_j = \mathcal{C}_{t} \setminus \Phi_j(\mathcal{C}_t)  \cup \{\emptyset\}$ is the set of coalitions outside $\Phi_j(\mathcal{C}_t)$;\strut}
%\State Find acceptable coalitions $$\quad~~\mathcal{D}_j = \br{\mathcal{S} \in \mathcal{C}_{t} \cup \{\emptyset\} \mid U_j(\mathcal{C}^{\mathcal{S}}_{t}) > U_j(\mathcal{C}_{t}), \mathcal{C}^{\mathcal{S}}_{t}\overset{j}{\longleftarrow} \mathcal{C}_t};$$
%\ForAll{$\mathcal{S} \in \Psi_j(\mathcal{C}_{t},\eta_j)$}
\ForAll{$\mathcal{S} \in \mathcal{C}_{t}$ s.t. $\tilde{U}_j(\mathcal{C}^{\mathcal{S}}_{t}, \eta_j) > \tilde{U}_j(\mathcal{C}_{t},\eta_j)$}
\State Ask members of $\mathcal{S}$ for permission to join;
\State Increment searching factor $\eta_j = \eta_j + 1$;
%\If {$U_k(\mathcal{C}^{\mathcal{S}}_{t}) \geq U_k(\mathcal{C}_{t}), \forall k \in \mathcal{S}$ and $\eta_j  \leq q_j$} 
\If{deviation $\mathcal{C}^{\mathcal{S}}_{t}\overset{j}{\longleftarrow} \mathcal{C}_t$ is admissible}
\State Leave $\Phi_j(\mathcal{C}_t)$ and join $\mathcal{S}$;
\State Update $\mathcal{C}_{t+1} = \mathcal{C}^{\mathcal{S}}_{t}$;
\State $t = t+1$;
\State Go to Line 2;
\EndIf
\EndFor
\EndFor
\Until No cell deviates
%\Statex \textbf{Output}: {$\mathcal{C}_t$}
\end{algorithmic} 
\end{algorithm}%
%
%\begin{definition}[Nash stability]\label{def:Nash_stability}
%A coalition structure $\mathcal{C}$ is Nash stable if there do not exist $j \in \mathcal{L}$ in a coalition $\Phi_j(\mathcal{C})$ and a coalition $\mathcal{S} \in \mathcal{C} \cup \{\emptyset\}$ such that $\mathcal{C}^{\mathcal{S}}_0 \overset{j}{\longleftarrow} \mathcal{C}_0$ satisfies $U_j(\mathcal{C}^{\mathcal{S}_0}) > U_j(\mathcal{C}_0)$.
%\end{definition}
%
%In a Nash stable coalition structure, no cell can strictly improve its performance by leaving its current coalition to join another coalition. Note that since a user knows the propagation parameters from all the other cells, a cell $j$ can calculate its utility function $U_j(\mathcal{C}_0)$ in~\eqref{eq:utility-functions} for its all possible deviations locally. 
%
\section{Simulations}\label{sec:simulations}%
In the simulations, we consider $S = 400$ transmission symbols (e.g., $T_c \!=\!4$ ms and $W_c\!=\!100$ kHz), SNR = $\frac{\rho}{\sigma^2}$ $=5$ dB, and a pathloss exponent of $3$. Each BS owns $10$ pilot sequences that are orthogonal to the pilots of all other BSs. Accordingly, we set the total number of pilots as $B = 10 L$ with $L$ being the number of BSs. For coalition formation, the searching budget of BS $j$ is set to $q_j=100$ for all $j\in \mathcal{L}$. We obtain the average performance from $10^3$ uniformly random BS deployments with uniform user distributions in each cell and a wrap-around topology, as exemplified in Fig.~\ref{fig:systemmodel}. 

\begin{figure}[t]
  \centering
  \includegraphics[width=\linewidth,clip]{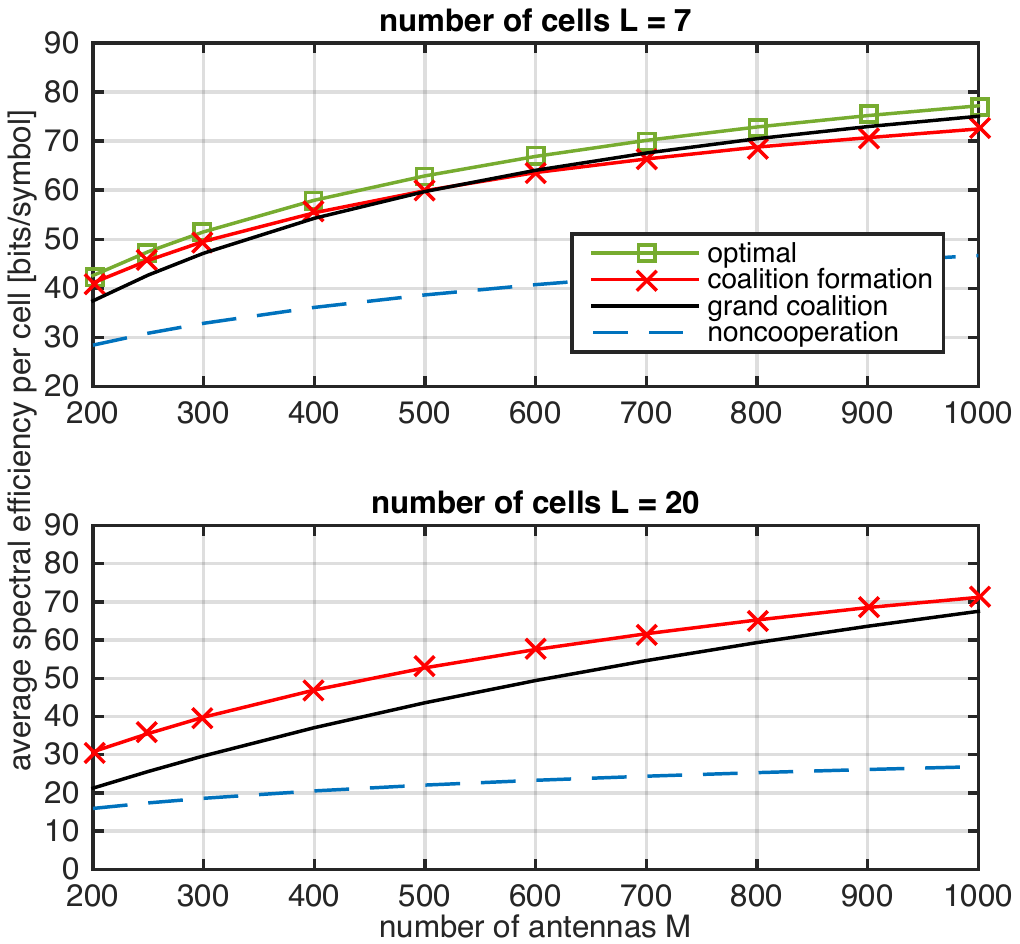}
  \caption{\label{fig:SE_M_MRC} Average spectral efficiency with MRC at the BSs.} 
  %\vspace{-6mm}
\end{figure}

\begin{figure}[t]
  \centering
  \includegraphics[width=\linewidth,clip]{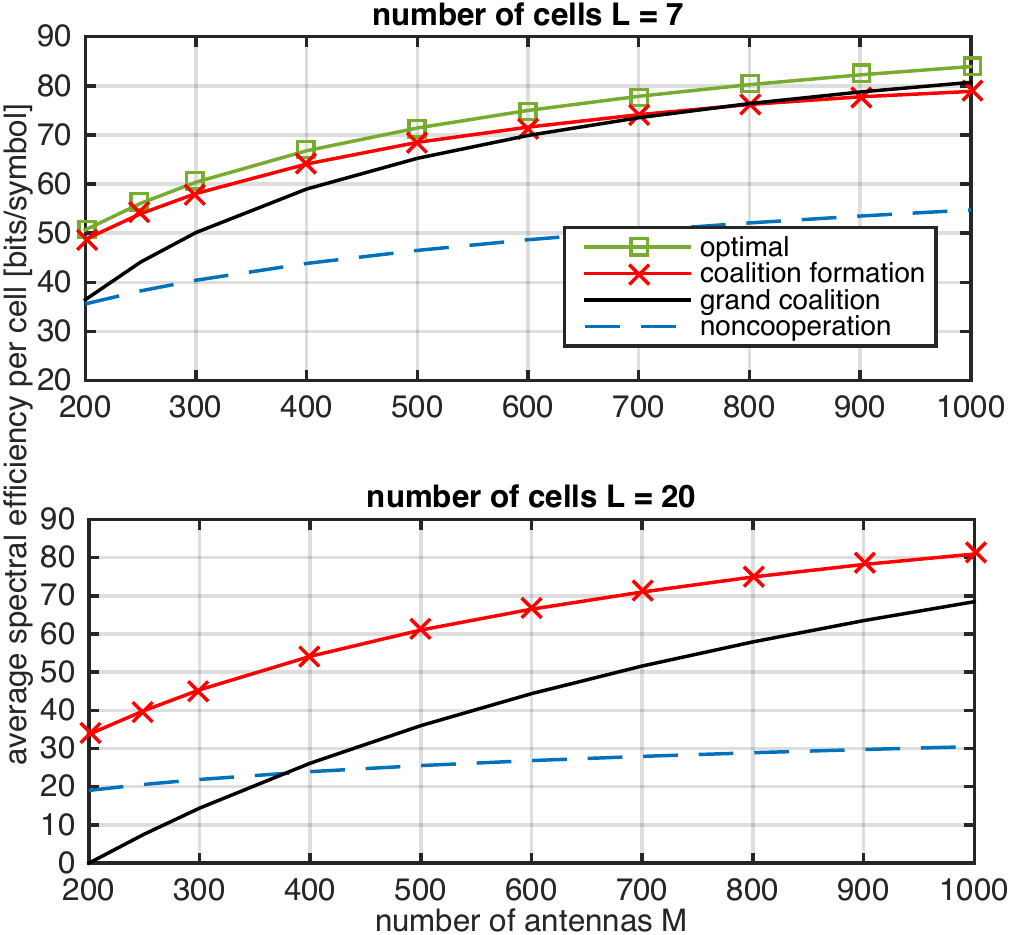}
  \caption{\label{fig:SE_M_ZFC} Average spectral efficiency with ZFC at the BSs.}
\end{figure}

In \figurename~\ref{fig:SE_M_MRC}, the average SE with MRC is plotted for different number of antennas at the BSs. Two scenarios are selected for which we ensure the same BS density of $25$ BSs per km$^2$ by appropriately choosing the region area the cells are deployed in. For $L=7$, finding the optimal coalition structure by exhaustive search is computationally possible. The other scenario is for $L=20$ BSs. The associated performance using the ZFC scheme in the same scenarios is provided in \figurename~\ref{fig:SE_M_ZFC}.

From \figurename~\ref{fig:SE_M_MRC}, the performance of coalition formation and the grand coalition (i.e., all BSs use all  pilot resources) can be observed to be close to optimal in the case of $7$ cells. The grand coalition outperforms the proposed coalition formation in the case of $7$ cells when $M>600$. Though the practical range of $M$ is less than $500$, the intersection point of the grand coalition and coalition formation curves generally depends on the choice of $B$. For fixed $B$, the high performance of the grand coalition with very large $M$ can be explained through the vanishing interference terms in the utility functions in Theorem~\ref{theorem:sum-spectral-efficiency}. Consequently, it is efficient to schedule as many UEs as possible in all cells, achieved by the grand coalition, in order to maximize the pre-log term in the average SE expression when $M$ is very large. In the ZFC scheme in \figurename~\ref{fig:SE_M_ZFC} for $L=7$, the grand coalition is more efficient than the proposed coalition formation when  $M >800$ for similar reasons. Here, it is observed that noncooperation outperforms the grand coalition with ZFC when the available degrees of freedom are small. In the case of $20$ cells in which $B = 200$, coalition formation provides performance gains compared to the other schemes in the selected range of $M$.

\begin{figure}[t]
  \centering
  \includegraphics[width=\linewidth,clip]{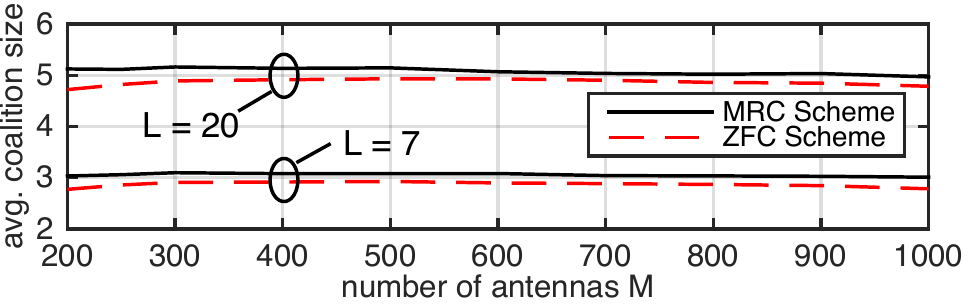}
  \caption{\label{fig:SIZE_M}Average coalition size after coalition formation.}
  \vspace{0.4cm}
  \includegraphics[width=\linewidth,clip]{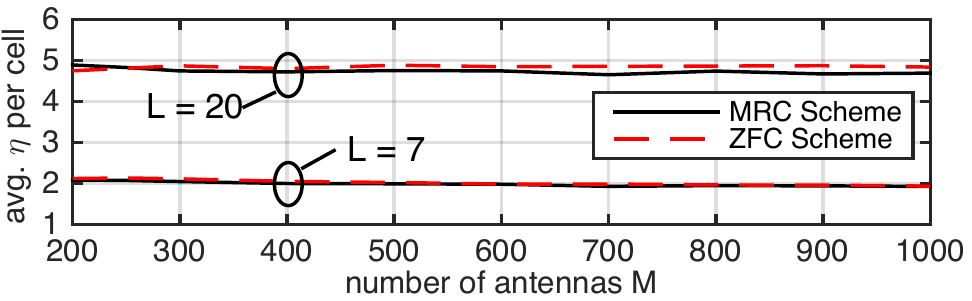}
  \caption{\label{fig:GAMMA_M}Average number of searches $\eta$ per BS.}
  \vspace{-3mm}
\end{figure}

The corresponding average coalition sizes for the plots in \figurename~\ref{fig:SE_M_MRC} and \figurename~\ref{fig:SE_M_ZFC} are shown in \figurename~\ref{fig:SIZE_M}. Similar average coalition sizes can be observed for the same number of BSs for MRC and ZFC. Evidently, the grand coalition is rarely reached with coalition formation. The complexity of coalition formation is reflected by the average number of searches by the BSs illustrated in \figurename~\ref{fig:GAMMA_M}. This number is incremented in Line 5 in Algorithm~\ref{alg:coalition0}. Although, we set the searching budget to $100$ in the simulations, the average number of searches is very low and about a quarter of the number of BSs. This measure does not show significant dependence on the number of antennas.%
\section{Conclusion}\label{sec:conclusion}

\vspace{-1mm}

A distributed algorithm was proposed for pilot allocation in the uplink of cellular massive MIMO networks of arbitrary shape. By assuming that each cell has a few unique pilots and can form coalitions with other cells to share pilot resources, a coalitional game in partition form was formulated. Each BS wants to maximize the average SE in its cell, taking pilots, CSI quality, and interference into account. The proposed mechanism has low complexity and provides performance gains compared to one-cell coalitions and full pilot reuse schemes. The solution is applicable also in the downlink, by capitalizing on uplink-downlink duality \cite{Bjornson2016a}. Future work will further analyze the performance dependencies on the total pilot budget as well as on the number of cells and their geometries.

%In the uplink of a cellular massive MIMO network, a distributed coalition formation algorithm is proposed to determine the pilot allocation across the cells. For this purpose, a coalitional game in partition form has been suitably applied in the setting in which a cell's average SE depends on the overall network coalition structure affecting both the CSI quality and interference. The proposed mechanism has low complexity and provides performance gains compared to one-cell coalitions and full pilot reuse schemes. Future work will further analyze the performance dependencies on the pilot budget per cell as well as on the number of cells and their geometries.
%\input{sections/appendix}
%
%
%\vspace{-1mm}

\bibliographystyle{IEEEtran}
\bibliography{IEEEabrv,refs}
\end{document}